\documentclass[12pt]{article}

\def\com#1#2{\bigl[#1,\,#2\bigr]}
\def\anticom#1#2{\bigl\{#1,\,#2\bigr\}}

\usepackage{amsmath,amssymb} 

\newcommand{\real}{\mathbb{R}}
\newcommand{\cS}{\mathcal{S}} 
\newcommand{\Psiv}{\varPsi}

\begin{document}

\title{\bf Supersymmetric Quantum Mechanics \\ of Scattering} 

\author{Toshiki Shimbori$^*$ and Tsunehiro Kobayashi$^\dag$ \\
{\footnotesize\it $^*$Institute of Physics, University of Tsukuba}\\
{\footnotesize\it Ibaraki 305-8571, Japan}\\
{\footnotesize\it $^\dag$Department of General Education 
for the Hearing Impaired,}
{\footnotesize\it Tsukuba College of Technology}\\
{\footnotesize\it Ibaraki 305-0005, Japan}}

\date{}

\maketitle

\begin{abstract}
 In the quantum mechanics of collision problems 
 we must consider scattering states of the system. 
 For these states, the wave functions 
 do not remain in Hilbert space, 
 but they are expressible 
 in terms of 
 generalized functions of a Gel'fand triplet. 
 Supersymmetric quantum mechanics 
 for dealing with the scattering states 
 is here proposed. 
\end{abstract}

\thispagestyle{empty}

\setcounter{page}{0}

\pagebreak

 SUSY (supersymmetry) is a concept 
 which connects between bosons and fermions~\cite{wb,west}. 
 SUSY QM (supersymmetric quantum mechanics) 
 is the more elementary concept, 
 because it forms a corner-stone 
 in the theory of SUSY in high-energy physics. 
 An example of a dynamical system in SUSY QM is 
 the SUSY HO (supersymmetric harmonic oscillator)~\cite{witten,cf}. 
 This example is of a stable system, so 
 the eigenvalue problem is solved in Hilbert space. 
 (For a review of SUSY QM, 
 see for example, reference~\cite{cks}.) 
 
 In the present paper 
 a radically different theory for SUSY is put forward, 
 which is concerned with collision problems in SUSY QM. 
 A simple and interesting model of resonance scattering 
 in quantum mechanics is 
 the PPB (parabolic potential barrier)~\cite{barton,
 bcd,bv,cdlp,sk,s2}. 
 This model is of importance for general theory, 
 because the eigenvalue problem of the PPB can be solved exactly 
 by a operator method on the same lines as 
 one has used for the HO~\cite{dirac,jjs}. 
 We must therefore begin to investigate the SUSY PPB in order to 
 set up the theoretical scheme 
 for dealing with the scattering states 
 of collision problems in SUSY QM. 
 
 \paragraph{The parabolic potential barrier \\}
 Let us first deal with a PPB~\cite{s2}, 
 a different problem from SUSY problem. 
 The Hamiltonian of the PPB is 
 \begin{equation}
  \hat{H}_b =\frac{1}{2m}\hat{p}^2 -\frac{1}{2} m\gamma^2 \hat{x}^2, 
   \label{2v.2.1} 
 \end{equation}
 where $m>0$ and also $\gamma>0$. 
 The standard states of this PPB have the wave functions, 
 which we may call $u^\pm_0$: 
 \begin{equation}
  u^\pm_0(x)= e^{\pm im\gamma x^2/2\hslash}. \label{2v.2.14}
 \end{equation}
 These $u^\pm_0$ do not belong to a Lebesgue space $L^2(\real)$, 
 but they are generalized functions 
 in ${\cS(\real)}^\times$ of 
 the following Gel'fand triplet~\cite{sk}: 
 \begin{equation}
  \cS(\real)\subset L^2(\real)\subset{\cS(\real)}^\times, 
   \label{2.2.15}
 \end{equation}
 where $\cS(\real)$ is a Schwartz space. 
 
 Introduce the normal coordinates~\cite{barton,bv,cdlp} 
 \begin{equation}
  \hat{b}^\pm\equiv\frac{1}{\sqrt{2m\hslash\gamma}}
   \left(\hat{p}\pm m\gamma\hat{x}\right), \label{2w.2.5} 
 \end{equation}
 essentially self-adjoint on $\cS(\real)$. 
 Using the commutation relation 
 $\com{\hat{x}}{\hat{p}}\equiv\hat{x}\hat{p}-\hat{p}\hat{x}=i\hslash$, 
 we find 
 \begin{equation}
  \left.
   \begin{gathered} 
    \com{\hat{b}^+}{\hat{b}^-}=i, \\ 
    \com{\hat{b}^+}{\hat{b}^+}=0,\,\,\,\com{\hat{b}^-}{\hat{b}^-}=0. 
   \end{gathered} \right\} \label{2w.2.7} 
 \end{equation}
 It should be noted that 
 the ambiguity of sign in the first of equations \eqref{2w.2.7} 
 is connected with 
 the choice of the arbitrary signs in \eqref{2w.2.5} 
 (cf. reference~\cite{s2}, equations (5) and (7)). 
 We also find that the Hamiltonian \eqref{2v.2.1} is 
 \begin{equation}
  \hat{H}_b=\frac{1}{2}\hslash\gamma\anticom{\hat{b}^+}{\hat{b}^-}, 
   \label{2w.2.8} 
 \end{equation}
 where 
 $\anticom{\hat{b}^+}{\hat{b}^-}
 \equiv\hat{b}^+\hat{b}^- +\hat{b}^-\hat{b}^+$. 
 Note that the extensions $\bigl(\hat{b}^\pm\bigr)^\times$ 
 of the normal coordinates 
 operating to a generalized function in ${\cS(\real)}^\times$ 
 have the meaning of $\hat{b}^\pm$ operating~\cite{s2}. 
 
 Let us now take $u^\pm_0$ and operate on them with $\hat{b}^\mp$. 
 Since we have $\hat{x}=x$ and $\hat{p}=-i\hslash\, d/dx$ 
 in the Schr\"{o}dinger representation, we obtain 
 \begin{equation}
  \hat{b}^\mp u^\pm_0(x) 
   =\frac{1}{\sqrt{2m\hslash\gamma}} 
   \left(-i\hslash\frac{d}{dx}\mp m\gamma x\right) 
   e^{\pm im\gamma x^2/2\hslash} =0, \label{2w.2.12}
 \end{equation}
 so that $\hat{b}^\mp$  applied to $u^\pm_0$ give zero. 
 Then 
 \begin{equation}
  \hat{H}_b u^\pm_0 =\frac{1}{2}\hslash\gamma\hat{b}^\mp\hat{b}^\pm u^\pm_0
   =\mp\frac{i}{2}\hslash\gamma u^\pm_0 
   \label{2w.2.17}
 \end{equation}
 and $u^\pm_0$ are generalized eigenstates 
 of $\hat{H}_b$ belonging to 
 the complex energy eigenvalues $\mp i\hslash\gamma/2$. 
 Similarly, 
 \begin{equation}
  \hat{H}_b \bigl(\hat{b}^\pm\bigr)^{n_b} u^\pm_0 
   =\mp i\left( n_b+\frac{1}{2}\right)\hslash\gamma
   \bigl(\hat{b}^\pm\bigr)^{n_b} u^\pm_0 \,\,\, 
   \left(n_b=0, 1, 2, \cdots\right). \label{2v.2.20} 
 \end{equation}
 Thus the states $\bigl(\hat{b}^\pm\bigr)^{n_b} u^\pm_0$ are 
 generalized eigenstates of $\hat{H}_b$ belonging to 
 the complex energy eigenvalues $\mp i(n_b+1/2)\hslash\gamma$. 
 
 Take this same Hamiltonian and apply it in the Heisenberg picture. 
 The Heisenberg equations of motion give 
 \begin{equation}
  \frac{d}{dt}\hat{b}^\pm(t) 
   =\frac{1}{i\hslash}\com{\hat{b}^\pm(t)}{\hat{H}_b} 
   =\pm\gamma\hat{b}^\pm(t), \label{2v.3.1} 
 \end{equation}
 and the solutions are 
 \begin{equation}
  \hat{b}^\pm(t)=\hat{b}^\pm e^{\pm\gamma t}. \label{2.3.3}
 \end{equation}
 The time factors of 
 \eqref{2.3.3} are the same as in the classical theory. 
 
 Again, we introduce some 
 essentially self-adjoint operators $\hat{d}^+$, $\hat{d}^-$ 
 to satisfy~\cite{s2} 
 \begin{equation}
  \left.
   \begin{gathered}
    \anticom{\hat{d}^+}{\hat{d}^-}=1,\\
    \anticom{\hat{d}^+}{\hat{d}^+}=0,\,\,\,
    \anticom{\hat{d}^-}{\hat{d}^-}=0,  
   \end{gathered} \right\} \tag{\ref{2w.2.7}$'$}\label{2.4.1}
 \end{equation}
 which are relations of the same form as \eqref{2w.2.7} 
 except for the anticommutators now replacing 
 the commutators there 
 and which therefore contain the ambiguity of sign. 
 Instead of \eqref{2w.2.8} we put 
 \begin{equation}
  \hat{H}_d =\hslash\gamma\hat{N}_d, 
   \tag{\ref{2w.2.8}$'$}\label{2x.4.9} 
 \end{equation}
 where $\hat{N}_d$ is the fermion number operator 
 \begin{equation}
  \hat{N}_d=-\frac{i}{2}\com{\hat{d}^+}{\hat{d}^-}. 
   \tag{\ref{2w.2.8}$''$}\label{2w.4.9} 
 \end{equation}
 
 We can treat the $\hat{d}^\pm$ 
 as we did the $\hat{b}^\pm$ in equations 
 \eqref{2w.2.12}, \eqref{2w.2.17} and \eqref{2v.2.20}. 
 We introduce the standard states $\phi^\pm_0$, satisfying 
 \begin{equation}
  \hat{d}^\mp \phi^\pm_0 =0, \tag{\ref{2w.2.12}$'$}\label{8.2.14} 
 \end{equation}
 and hence 
 \begin{equation}
  \hat{H}_d \phi^\pm_0 
   =\pm\frac{i}{2}\hslash\gamma\hat{d}^\mp\hat{d}^\pm \phi^\pm_0
   =\pm\frac{i}{2}\hslash\gamma \phi^\pm_0, 
   \tag{\ref{2w.2.17}$'$}\label{8w.2.15}
 \end{equation}
 like \eqref{2w.2.17}. 
 Again 
 $$
 \hat{H}_d \hat{d}^\pm\phi^\pm_0 
 =\mp\frac{i}{2}\hslash\gamma \hat{d}^\pm\phi^\pm_0, 
 $$ 
 showing that $\phi^\pm_1\equiv\hat{d}^\pm\phi^\pm_0$ and 
 $\phi^\mp_0$ are twofold degenerate states 
 belonging to the complex energy eigenvalues $\mp i\hslash\gamma/2$. 
 However, $\bigl(\hat{d}^\pm\bigr)^2 \phi^\pm_0$ are zero 
 from \eqref{2.4.1}. 
 Instead of \eqref{2v.2.20} we now have 
 \begin{equation}
  \hat{H}_d \bigl(\hat{d}^\pm\bigr)^{n_d} \phi^\pm_0 
   =\mp i\left( n_d-\frac{1}{2}\right)\hslash\gamma
   \bigl(\hat{d}^\pm\bigr)^{n_d} \phi^\pm_0 \,\,\, 
   \left(n_d=0, 1\right). 
   \tag{\ref{2v.2.20}$'$}\label{8.2.17} 
 \end{equation}
 
 In the Heisenberg picture, 
 equations \eqref{2v.3.1} still hold, 
 with $\hat{b}^\pm(t)$ replaced by $\hat{d}^\pm(t)$, so 
 \begin{equation}
  \hat{d}^\pm(t)=\hat{d}^\pm e^{\pm\gamma t}. 
   \tag{\ref{2.3.3}$'$}\label{8.2.19} 
 \end{equation}
 From this we see, 
 bearing in my mind the result \eqref{2.3.3}, 
 that the $\hat{d}^\pm(t)$ have just the same 
 time factors as the $\hat{b}^\pm(t)$. 
 
 \paragraph{The supersymmetric parabolic potential barrier \\}
 Let the SUSY Hamiltonian of the SUSY PPB be  
 \begin{equation}  
  \hat{H}=\hat{H}_b+\hat{H}_d, \label{8.3.1} 
 \end{equation}
 where $\hat{H}_b$ is given by \eqref{2v.2.1} or \eqref{2w.2.8} 
 and $\hat{H}_d$ is given by \eqref{2x.4.9} and \eqref{2w.4.9}. 
 Using the values of the commutators and anticommutators 
 given by \eqref{2w.2.7} and \eqref{2.4.1}, we get 
 $$
 \hat{H}
 =\hslash\gamma\bigl(\hat{b}^+\hat{b}^- -i\hat{d}^+\hat{d}^-\bigr)
 =\hslash\gamma\bigl(\hat{b}^-\hat{b}^+ +i\hat{d}^-\hat{d}^+\bigr), 
 $$
 and hence the SUSY Hamiltonian \eqref{8.3.1} 
 is essentially self-adjoint. 
 
 Let us consider the essentially self-adjoint operators 
 $\hat{Q}^+$, $\hat{Q}^-$ defined by 
 \begin{equation}
  \hat{Q}^\pm\equiv\sqrt{\hslash\gamma}\hat{b}^\mp\hat{d}^\pm. 
   \label{8.3.3}
 \end{equation}
 Since the time factors in formulas \eqref{2.3.3} and \eqref{8.2.19} 
 cancel out in their products of \eqref{8.3.3}, 
 they are constants of the motion. 
 This leads, as will be shown in equations \eqref{8.3.5}, 
 to the result that 
 $\hat{Q}^\pm$ are the {\it supercharges} of the SUSY PPB.
 We must evaluate the commutators and anticommutators of 
 the supercharges with 
 the normal coordinates $\hat{b}^\pm$, $\hat{d}^\pm$, 
 the SUSY Hamiltonian $\hat{H}$, and with each other. 
 Using the laws \eqref{2w.2.7} and \eqref{2.4.1}, we obtain 
 \begin{equation}
 \left.
  \begin{gathered} 
   \com{\hat{Q}^\pm}{\hat{b}^\pm}
   =\mp i\sqrt{\hslash\gamma}\hat{d^\pm},\,\,\, 
   \anticom{\hat{Q}^\pm}{\hat{d}^\mp}
   =\sqrt{\hslash\gamma}\hat{b^\mp}, \\
   \com{\hat{Q}^\pm}{\hat{b}^\mp}=0,\,\,\, 
   \anticom{\hat{Q}^\pm}{\hat{d}^\pm}=0, 
  \end{gathered} \right\} \label{8w.3.4}
 \end{equation}
 and similarly, 
 \begin{equation}
  \com{\hat{Q}^\pm}{\hat{H}}=0. \label{8.3.5}
 \end{equation}
 Again 
 \begin{equation}
  \left.
   \begin{gathered} 
    \anticom{\hat{Q}^+}{\hat{Q}^-}=\hat{H}, \\
    \anticom{\hat{Q}^+}{\hat{Q}^+}=0,\,\,\, 
    \anticom{\hat{Q}^-}{\hat{Q}^-}=0. 
   \end{gathered} \right\} \label{8.3.6} 
 \end{equation}
 Equations \eqref{8w.3.4} show that 
 $\hat{Q}^\pm$ make the {\it SUSY transformation} 
 which interchanges the operators of 
 ``bosonic'' and ``fermionic''. 
 We have in \eqref{8.3.6} 
 the {\it SUSY algebra} in the SUSY PPB. 
 The first of equations \eqref{8.3.6}, however, 
 does not mean that 
 the SUSY Hamiltonian $\hat{H}$ is positive definite. 
 
 We can form 
 the generalized Fock spaces of the SUSY PPB 
 on the same lines as the SUSY HO~\cite{witten,cf,cks}. 
 We now consider the following states: 
 \begin{equation}
  \Psiv^{\pm\pm}_{n_b n_d}=
   \bigl(\hat{b}^\pm\bigr)^{n_b}\bigl(\hat{d}^\pm\bigr)^{n_d} 
   u^\pm_0 \phi^\pm_0. \label{8.3.7}
 \end{equation}
 The right-hand sides here are undetermined to the extent of 
 arbitrary numerical factors. 
 We may consider the states $\Psiv^{\pm\pm}_{0 0}$ as standard states, 
 since 
 \begin{equation}
  \hat{H} \Psiv^{\pm\pm}_{0 0}=0, \label{8.3.8}
 \end{equation}
 {\it both states $\Psiv^{\pm\pm}_{0 0}$ have zero energy eigenvalue}. 
 Also 
 \begin{equation}
  \hat{Q}^+ \Psiv^{\pm\pm}_{0 0}=\hat{Q}^- \Psiv^{\pm\pm}_{0 0}=0. 
   \label{8.3.9}
 \end{equation}
 This shows that 
 {\it the states $\Psiv^{\pm\pm}_{0 0}$ are supersymmetrical}. 
 Thus the standard states $\Psiv^{++}_{0 0}$, $\Psiv^{--}_{0 0}$
 are twofold degenerate. 
 Further, if we shall consider 
 the degree of degeneracy of the fermion sector 
 (caused by the doublets $\left(\phi^+_0, \phi^-_1\right)$ and 
 $\left(\phi^-_0, \phi^+_1\right)$), 
 we now have the SUSY-quartet 
 consisting of four kinds of standard states, 
 $\left(\Psiv^{++}_{0 0}, \Psiv^{+-}_{0 1}, 
 \Psiv^{-+}_{0 1}, \Psiv^{--}_{0 0}\right)$. 
 It is interesting that 
 such a stable idea as 
 zero energy eigenvalue should appear in 
 the SUSY PPB in this way. 
 These stationary states of the SUSY PPB are analogous to 
 the stationary flows of the 2D PPB~\cite{sk4}. 
 The energy eigenvalues of the other states 
 can be obtained from \eqref{8.3.7}. 
 We have from \eqref{2v.2.20} and \eqref{8.2.17} 
 \begin{gather}
  \hat{H} \Psiv^{\pm\pm}_{n_b n_d}=E^{\pm\pm}_{n_b n_d} 
  \Psiv^{\pm\pm}_{n_b n_d}, \label{8.3.10} 
  \intertext{where} 
  E^{\pm\pm}_{n_b n_d}=\mp i\left(n_b+n_d\right)\hslash\gamma \,\,\, 
  \left(n_b=0, 1, 2,\cdots, \text{ and } n_d=0, 1\right). \label{8.3.11}
 \end{gather}
 Thus the states $\Psiv^{\pm\pm}_{n\, 1}$ and $\Psiv^{\pm\pm}_{n+1\, 0}$ 
 are eigenstates of $\hat{H}$ belonging to 
 the same complex energy eigenvalues $\mp i(n+1)\hslash\gamma$ 
 with $n=0, 1, 2, \cdots$, respectively. 
 This result may be verified by \eqref{8.3.5}, 
 since, by applying the supercharges $\hat{Q}^\pm$ to these states, 
 we can get 
 $$
 \hat{Q}^\mp \Psiv^{\pm\pm}_{n\, 1}\propto\Psiv^{\pm\pm}_{n+1\, 0},\,\,\, 
 \hat{Q}^\pm \Psiv^{\pm\pm}_{n+1\, 0}\propto\Psiv^{\pm\pm}_{n\, 1}. 
 $$
 Provided that 
 we take account of the twofold degeneracy of the fermion sector, 
 $\left(\Psiv^{++}_{n\, 1}, \Psiv^{+-}_{n\, 0},\right.$ 
 $\left.\Psiv^{+-}_{n+1\, 1}, \Psiv^{++}_{n+1\, 0}\right)$ and 
 $\left(\Psiv^{--}_{n\, 1}, \Psiv^{-+}_{n\, 0}, 
 \Psiv^{-+}_{n+1\, 1}, \Psiv^{--}_{n+1\, 0}\right)$ 
 will in general form quartets for the SUSY PPB. 
 
 \paragraph{The superpotential \\}
 The above analysis can be extended to the SUSY problem of 
 scattering. 
 We introduce an arbitrary real function $W(x)$ 
 which satisfies, 
 as the generalization of \eqref{2w.2.5} and \eqref{8.3.3}, 
 \begin{equation}
  \hat{Q}^\pm 
   =\frac{1}{\sqrt{2m}}\left[\hat{p}\mp W(x)\right]\hat{d}^\pm. 
   \label{8w.4.1}
 \end{equation}
 We call $W(x)$ the {\it superpotential} 
 for the scattering process in SUSY QM, 
 to keep up the analogy with the usual formulation 
 of SUSY QM~\cite{witten,cf,cks}. 
 
 The SUSY Hamiltonian for the scattering process is, 
 from the first of equations \eqref{8.3.6} 
 which are valid also for the general theory, 
 \begin{equation}
  \hat{H}=\frac{1}{2m}\left[\hat{p}^2 -W(x)^2\right]
   +\frac{\hslash}{m} W(x)^\prime\hat{N}_d, \label{8w.4.2} 
 \end{equation}
 with $\hat{N}_d$ given by \eqref{2w.4.9}. 
 Note that the second term in the $[$ $]$ brackets in \eqref{8w.4.2}, 
 which is the part of $\hat{H}$ referring to 
 the potential energy for the scattering process, 
 appears with a minus sign. 
 One can check that $\hat{Q}^\pm$ commute with $\hat{H}$ 
 and are constants of the motion. 
 
 Let us write the standard states which are supersymmetrical. 
 These states will correspond to 
 wave functions $\Psiv^{\pm\pm}_{0 0}$, say, satisfying 
 \begin{equation}
  \hat{Q}^\mp\Psiv^{\pm\pm}_{0 0}=0,\,\,\, 
   \hat{Q}^\pm\Psiv^{\pm\pm}_{0 0}=0. \label{8.4.3}
 \end{equation}
 The first of these equations tells us that 
 the wave functions will be of the form 
 \begin{equation}
  \Psiv^{\pm\pm}_{0 0}(x)=u^\pm_0(x) \phi^\pm_0, 
   \label{8.4.4} 
 \end{equation}
 where $\phi^\pm_0$ satisfy \eqref{8.2.14}. 
 With the help of this result 
 the second of equations \eqref{8.4.3}, 
 written in terms of $x$-representatives, becomes 
 \begin{equation}
  \left[-i\hslash\frac{d}{dx}\mp W(x)\right] u^\pm_0(x)=0. 
   \label{8w.4.5}
 \end{equation}
 Hence we get 
 \begin{equation}
  u^\pm_0(x)=
   \exp\left[\pm\frac{i}{\hslash}\int^x W(x^\prime) dx^\prime\right], 
   \label{8.4.6}
 \end{equation}
 except for the numerical factors. 
 
 Our work on the SUSY PPB in equations \eqref{8.3.1}--\eqref{8.3.11} 
 provides an example of a superpotential of resonance scattering. 
 Equation \eqref{8.3.1} is of the form \eqref{8w.4.2} 
 with $m\gamma x$ for $W(x)$, and it shows that 
 the wave functions \eqref{8.4.6} agree with \eqref{2v.2.14}. 
 It should be noted that 
 $u^+_0$ represents particles moving outward to 
 the infinity $|x|=\infty$, and 
 $u^-_0$ represents particles moving inward to 
 the origin $x=0$~\cite{sk}. 
 The result for the SUSY PPB is still valid 
 when the superpotential $W(x)$ is an odd function of $x$. 
 
 On the other hand, 
 when the superpotential $W(x)$ is an even function of $x$, 
 the behaviors of the standard states may be changed. 
 Let us see 
 what the above results become in the simple case 
 when $W(x)=p>0$ (a real constant). 
 Equations \eqref{8.4.6} for this case read 
 $$u^\pm_0(x)=e^{\pm ipx/\hslash}\in{\cS(\real)}^\times, $$ 
 showing that $u^+_0$ describes plane waves 
 moving to the $+x$-direction, 
 and $u^-_0$ describes plane waves 
 moving to the $-x$-direction. 
 The result for the SUSY free particle is valid 
 whenever the superpotential $W(x)$ is an even function of $x$. 
 
 The theory that has been set up here is 
 applicable to collision problems in SUSY QM. 
 If we take the usual formulation of SUSY QM~\cite{witten,cf,cks}, 
 we may set up the above-mentioned scheme 
 by taking the superpotential, $W(x)$, 
 in $N=2$ SUSY QM, 
 and replacing it 
 by the method of complex scaling~\cite{moiseyev}. 
 
 \pagebreak

\end{document}